\documentclass[12pt,epsf]{article}

\newlength{\pubnumber} 

\catcode`\@=11
\@addtoreset{equation}{section}

\def\section{\@startsection{section}{1}{\z@}{3.5ex plus 1ex minus .2ex}
 {2.3ex plus .2ex}{\large\bf}}
\def\subsection{\@startsection{subsection}{2}{\z@}{2.3ex plus .2ex}
 {2.3ex plus .2ex}{\bf}}

\usepackage{graphicx}% Include figure files
\usepackage{dcolumn}% Align table columns on decimal point
\usepackage{bm}% bold math

\usepackage[margin=1in]{geometry}

\usepackage{amsmath}
\usepackage{amsfonts}
\usepackage{amssymb}
\usepackage{hyperref}

\def\beq{\begin{equation}}
\def\eeq{\end{equation}}
\def\beqn{\begin{eqnarray}}
\def\eeqn{\end{eqnarray}}

\def\nolabel{\nonumber }

\def\mod{{\rm mod\ }}

\def\nolabel{\nonumber }

\begin{document}

\begin{titlepage}
\setcounter{page}{1}
\rightline{BU-HEPP-08-18}
\rightline{CASPER-08-07}
\rightline{\tt }

\vspace{.06in}
\begin{center}
{\Large \bf Grover's Quantum Search Algorithm and\\ Free Fermionic Heterotic Models}
\vspace{.12in}

{\large
        Matthew B. Robinson\footnote{m\_robinson@baylor.edu} and
        Gerald B. Cleaver\footnote{gerald\_cleaver@baylor.edu}}
\\
\vspace{.12in}
{\it Center for Astrophysics, Space Physics \& Engineering Research\\
     Department of Physics, One Bear Place \# 97316\\
     Baylor University\\
     Waco, TX 76798-7316\\}
\vspace{.06in}
\end{center}

\begin{abstract}
Given an efficient and systematic method for generating input sets for free fermionic heterotic model building we consider what the realistic bounds are for a statistical analysis of the free fermionic Landscape with a classical computer.  We then consider what kind of improvement could be expected on a quantum computer.  We do outline the basic structure of the relevant quantum algorithms, but we do not detail the construction of the oracle which would be involved in the calculations.  
\end{abstract}

\end{titlepage}
\setcounter{footnote}{0}

%***********************************************************************************************

\section{Free Fermionic Models and the String Landscape}\label{sec:ffm1}

During the decade following the first string revolution, the focus of string phenomenology was two-fold: 
development of methods for constructing consistent string models with compactified dimensions and searching within the domain of each construction method for those models with at least {\it quasi-realistic} phenomenology \cite{methods}. Many three generation string models were found within a few years. The dominant view back then was that discovery of the true string vacuum (model) was within reach--that the string ``needle" would eventually be found within the stringstack of but a few trillion vacua. Eventually a handful of quite realistic MSSM or Near-MMSM three generation models were indeed found \cite{mssmbook}, especially following the first {\it Minimal Standard Heterotic String Model} \cite{mshsm}. 

However, following the second string revolution the rise of $M$-theory has taught string phenomenologists the likely impossibility of finding a ``true" string vacuum somewhere on the string/M-landscape composed of {\it at least} $10^{500}$ vacua. In addition to the vast number of vacua, all vacua now appear to be on equal footing. Thus, the phenomenological goal has shifted from studying individual string models to better understanding statistically the characteristics of the string/M-models on the landscape, or at least within specific domains \cite{statstudies}. 

The (often overlapping) domains on the landscape frequently correspond to model construction methods. One construction method that has been widely explored in terms of individual models and for which large scale statistical studies are underway is free fermions \cite{fff1,fff2}.
The free fermionic heterotic string has provided many quasi-realistic (Near-)MSSM-like models \cite{nmssm,freegen,optun}, semi-GUT models \cite{sguts,psm2}, and GUT models \cite{fsu5}. 
In the context of the second string revolution, the important features to now investigate are the overall phenemological patterns within these and the many more, as yet undiscovered, free fermionic models. Several studies of the 
statistical properties and characteristics of free fermionic heterotic models are underway by different research groups \cite{af1,kd,mr1,ro1,mr2,jg1}. 

An algorithm to systematically and efficiently generate free fermionic heterotic models was recently introduced \cite{mr1}. As a first application, we have initiated an indepth study of the statistics of the gauge groups in free fermionic heterotic strings. Our approach enables a {\it complete} study of all gauge group models 
to be generated and analyzed with extreme efficiency, up to continually increasing Layers (the number of gauge basis vectors) and Orders (the lowest positive integer multiple $N$ that transforms each basis vector back into the untwisted sector mod(2)). Our algorithm significantly increases by several orders of magnitude the efficiency of a systematic statistical study of regions of the parameter space of free fermionic heterotic models \cite{mr1}. 
Nevertheless, current computational capabilities pose a serious limit to the range of parameter spaces that can be systematically investigated.

Thus, given an efficient and systematic method for generating the required input data for a given string model construction method, such as the free fermionic heterotic approach, we consider what the realistic bounds are for a statistical analysis of the Landscape with a classical computer.  We then consider what kind of improvement could be expected on a quantum computer.  We do outline the basic structure of the relevant quantum algorithms, but we do not detail the construction of the oracle which would be involved in the calculations.  
Since we will be relating quantum computing techniques to free fermionic construction, we briefly review in Section Two free fermionic heterotic model building and modular invariance constraints. Then in Section Three we review the
general concepts of quantum computing. In Section Four we discuss Grover's Quantum Search Algorithm, followed in  
Section Five by its application to the statistics of the free fermionic Landscape. Section Six concludes our discussions.

\section{Free Fermionic Heterotic Models}\label{sec:ffhm}

The first object required to specify a model in the free fermionic heterotic string \cite{fff1,fff2}
is a set
\begin{eqnarray}
A = \{ \vec \alpha^i \in \mathbb{Q}^{64}\cap (-1,1]^{64}| i \in \{1,...,L\in \mathbb{N}\}\},
\end{eqnarray}
where components $\alpha^i_j$, $j=1,\ldots,20$ are boundary conditions for real worldsheet free fermion degrees of freedom of the left-moving supersymmetric string,
 and $\alpha^i_j$ for $j=21,\ldots,64$ are boundary conditions for real worldsheet free fermion degrees of freedom on the right moving bosonic string. In the $\mathbb{C}$ basis (complex fermions), each component of $\vec \alpha^i$ is double counted, and $\vec \alpha^i$ is a $32$ $(10+22)$ component vector (which can be generalized to include left-right paired real fermions).  

The order $N_i$ of a given $\vec \alpha^i$ is defined as\footnote{We include ``0" in the set of natural numbers $\mathbb{N}$ as in set theory, in contrast to its general exclusion in number theory.}
\begin{eqnarray}
N_i \equiv \min \{m\in \mathbb{N} \mid m \alpha^i_j = 0 \; \mod \, 2\;  \forall j\},
\end{eqnarray}
with
\begin{eqnarray}
N_{ij} \equiv {\rm LCM}(N_i,N_j).
\end{eqnarray}
Thus, each component of an $\vec \alpha$ of order $N$ is of the form
\begin{eqnarray}
(\vec \alpha)_i \in {2\mathbb{Z}\over N} \cap [-1,1) \label{eq:elementsofalpha}
\end{eqnarray}

Modular Invariance demands 
\begin{eqnarray}
&&N_i \; \vec \alpha^i \cdot \vec \alpha^i = 0 \; \mod \, 8 \label{eq:modinv1}\\
&&N_{ij} \; \vec \alpha^i \cdot \vec \alpha^j = 0 \; \mod \, 4 \label{eq:modinv2}
\end{eqnarray}
in the $\mathbb{C}$ basis \cite{fff1,fff2},
along with requiring that the number of real rermions simultaneously periodic for any three
basis vectors is even (including cases where one or more of the basis vectors may be the same).
Further, each model contain $\vec \alpha^1 = \mathbb{I}$, the 64 real-component vector with every element 
equal to one.  

In a given free fermionic model, the different sectors of a model are formed by all linear combinations of the $\vec \alpha^i$'s, with coefficients $m^k \in \mathbb{N}$ where each coefficient $m^k_i < N_i$.  Each linear combination, or sector, is denoted 
\begin{eqnarray}
\vec V^k = \sum_{i=1}^{L} m^k_i \; \vec \alpha^i \; \mod \, 2.\label{eq:defV}
\end{eqnarray}
We can (and will) think of each set of coefficients $m^k$ as an L-dimensional vector in $\mathbb{N}^L$, whose $i^{th}$ component is constrained by the order of $\vec \alpha^i$.

For a given sector $\vec V^k$, a worldsheet fermion $f_j$ transforms as 
\begin{eqnarray}
f_j \rightarrow -e^{i \pi V^k_j}f_j
\end{eqnarray}
around non-contractible loops on the worldsheet.  Thus, for $\mathbb{R}$ fermions, $V^k_j$ must be either 0 or 1, whereas for $\mathbb{C}$ fermions, $V^k_j$ must be rational.  
For each sector, we can form the $U(1)$ charges for the Cartan generators of the unbroken gauge groups (which are in one to one correspondence with the $U(1)$ currents $f^*_jf_j$ for each complex fermion $f_j$);
\begin{eqnarray}
\vec Q_{\vec V^k} \equiv {1 \over 2} \vec V^k + \vec F^k \label{eq:defQ}
\end{eqnarray}
where $\vec F^k$ is a fermion number operator which counts each mode of $f_j$ once and of $f^*_j$ minus once.  Or, in other words, $F^k_i \in \{-1,0,1\}\; \forall\; i$.

The second object required to specify a model in the free fermionic heterotic string is an $L \times L$ matrix $k_{ij}$.  Modular Invariance imposes the following constraints on  $k_{ij}$;
\begin{eqnarray}
&&k_{ij} + k_{ji} = {1\over2} \vec \alpha^i \cdot \vec \alpha^j \; \mod \, 2 \label{eq:kijone}\\
&&k_{ii}+k_{i1} = {1\over4} \vec \alpha^i \cdot \vec \alpha^i - s_i \; \mod \, 2 \label{eq:kijtwo}\\
&&N_j k_{ij} = 0 \; \mod \, 2,  \label{eq:kijthree}
\end{eqnarray}
where $s_i$ is the 4 dimensional spacetime component of $\vec \alpha^i$. 
Furthermore, $k_{ij} \in (-1,1]$.
The GSO projection constraint for physical states is;
\begin{eqnarray}
\vec \alpha^i \cdot \vec Q_{\vec V^k} = \sum_{n=1}^L m^k_n k_{in}+s_i \; \mod \, 2. \label{eq:gso}
\end{eqnarray}

This completes our review of the general construction the free fermionic heterotic string models. 
We now now consider how quantum computing could significantly aid a systematic statistical study of this region (or of any other region) of the Landscape.

\section{Quantum Computation}\label{sec:qc1} 

In the years since 1965 when Gordon Moore proposed it, Moore's Law has approximately held true.  However, due to the drastic decreases in processor size and the increasing effects of quantum mechanics, it appears that fundamental fabrication barriers lie in the not too distant future.\footnote{The content of this introduction comes primarily from \cite{quantcom} and \cite{mer}} The physical construction of classical computers seems to demand a ``smallest scale" beyond which we cannot go, and it is likely that we will reach this scale in the next decade or two.  

One solution to this problem is to reformulate the physical construction of computers.  One such paradigm is provided by Quantum Computation.  The basic idea of Quantum Computation is that that the smallest unit of information is not a bit, but rather is a quantum bit, or q-bit, which can exist in a superposition of $1$ and $0$ at the same time;
\begin{eqnarray}
|\Psi\rangle = \alpha |0\rangle + \beta|1\rangle, \nolabel
\end{eqnarray}
where $|\alpha|^2 + |\beta|^2 =1$, and $|1\rangle = \begin{pmatrix} 1 \\ 0 \end{pmatrix}$ and $\begin{pmatrix} 0 \\ 1 \end{pmatrix}$ are the computational basis vectors.  
 
There then exists a set of universal quantum gates which, similarly to the classical universal gates, can be used to build an arbitrary operation on a set of q-bits.  As an example of why such a system is useful, consider one such universal quantum gate, the Hadamard gate
\begin{eqnarray}
H = {1\over \sqrt{2}} \begin{pmatrix} 1 & 1 \\ 1 & -1\end{pmatrix} \label{eq:hadamard}
\end{eqnarray}
acting on each q-bit in an $n$ q-bit register $|0\rangle^{\otimes n}= |0\rangle_1 |0\rangle_2 \cdots |0\rangle_n$.  This will give
\begin{eqnarray}
H^{\otimes n} |0\rangle^{\otimes n} = {1\over \sqrt{2^n}} \sum_{i=0}^{2^n-1} |i\rangle^{\otimes n}, \label{eq:superposition}
\end{eqnarray}
where $i$ runs over all $2^n$ $n$ q-bit states.  With a classical computer, we could act on a single $n$ bit register with some operator $C$, taking $C|i\rangle^{\otimes n} \rightarrow |i'\rangle^{\otimes n}$, and this operator must be used once for every possible input.  However, with the quantum computer, we can act on the superposition (\ref{eq:superposition}) a \it single \rm time with some operator $U$ (we use $U$ because quantum operators must be Unitary), and 
\begin{eqnarray}
UH^{\otimes n} |0\rangle^{\otimes n} = {1 \over \sqrt{2^n}}\sum_{i=0}^{2^n-1} |i'\rangle^{\otimes n}
\end{eqnarray}
will be a superposition of every outcome at once.  So using only $n$ q-bits, and a single use of $U$, we have $2^n$ transformations.  

While this is impressive, there is a serious drawback.  While this paradigm offers a massive parallelism, we obviously don't have access to every state in the superposition.  All we can do is make a measurement of each q-bit, and we have equal probability of getting any $n$ q-bit state.  Getting a practical result out of a quantum computer therefore takes quite a bit more work.  We mention this example merely to give an idea of how quantum mechanics could be used in computation.  

Despite decades of work, there has been relatively little progress made in developing algorithms for quantum computers.  One problem is that a quantum algorithm is only acceptable if it offers a speedup over the fastest classical algorithms.  Another problem is that all of our intuitions come from the classical (looking) world we experience, and therefore we lack clarity in ``thinking quantum" in order to design algorithms.  However, there are a few quantum algorithms that have been developed and which offer a significant  increase in efficiency over their classical counterparts.  These existing algorithms could potentially be used in generating statistics on the free fermionic heterotic Landscape.  We will therefore outline how these algorithms work, and discuss what type of improvement they would offer in such an analysis.  
To date, there are no quantum computers in active operation.  However, it is likely that in the next few decades their production will allow for applications like the one discussed in this paper.

\section{Quantum Algorithms}\label{sec:qa1}

For a more detailed exposition of the nature of quantum gates and their physical construction, see \cite{quantcom}, which we follow closely in this section.  It should be assumed that all of the gates discussed below are realizable.  

\subsection{Quantum Fourier Transform}

The first quantum operation we consider is the quantum Fourier transform (QFT) \cite{quantcom}.  The action of the QFT on an $n$ q-bit state $|j\rangle^{\otimes n}$ is defined by
\begin{eqnarray}
|j\rangle^{\otimes n} \rightarrow {1\over \sqrt{2^n}} \sum_{k=0}^{2^n-1} e^{{(2\pi i) jk \over 2^n}}|k\rangle^{\otimes n}.
\end{eqnarray}
But we can re-express this as
\begin{eqnarray}
|j\rangle^{\otimes n} &\rightarrow& {1\over \sqrt{2^n}} \sum_{k=0}^{2^n-1} e^{{(2\pi i) jk \over 2^n}}|k\rangle^{\otimes n}\\
 &=& {\big(|0\rangle + e^{2\pi i 0.j_n}|1\rangle\big)\big(|0\rangle + e^{2\pi i 0.j_{n-1}j_n}\big)\cdots \big(|0\rangle + e^{2\pi i 0.j_1j_2 \cdots j_n}|1\rangle\big) \over \sqrt{2^n}}, \label{eq:qft}
\end{eqnarray}
where $0.j_kj_{k+1}\cdots j_m \equiv {j_k \over 2} + {j_{k+1} \over 4}+ \cdots {j_m \over 2^{m-k+1}}$ is the binary fraction for $j$, where each $j_i$ is defined by $j=j_12^{n-1}+j_22^{n-2}+\cdots + j_n2^0$.  

\subsection{Phase Estimation}\label{sec:phase}

Suppose we have some unitary operator $U$ with eigenvector $|u\rangle$ and eigenvector $e^{2\pi i \phi}$, where $\phi$ is unknown.  We can use the QFT (\ref{eq:qft}) to estimate $\phi$ to arbitrary accuracy \cite{quantcom}.  
We assume that we have oracles which will both prepare the state $|u\rangle$ as well as perform a controlled-$U^{2^j}$ operation for non-negative integers $j$.  
We then prepare a register with $n$ q-bits all in the state $|0\rangle$ ($|0\rangle^{\otimes n}$), where $n$ depends on how accurately we want to estimate $\phi$, and a second register in the state $|u\rangle$, or
\begin{eqnarray}
|0\rangle^{\otimes n} |u\rangle.
\end{eqnarray}

We act on each of the q-bits in the first register with a Hadamard gate (\ref{eq:hadamard}), creating the state
\begin{eqnarray}
{1\over \sqrt{2^n}} \big(|0\rangle + |1\rangle\big)\big(|0\rangle + |1\rangle\big)\cdots \big(|0\rangle + |1\rangle\big) \; |u\rangle.
\end{eqnarray}
Next, we act on the second register, $|u\rangle$, with a controlled-$U$ operation with $U$ raised to successive powers of $2$, creating the state
\begin{eqnarray}
& & {1\over \sqrt{2^n}} \big(|0\rangle +e^{2\pi i 2^{n-1}\phi} |1\rangle\big)\big(|0\rangle + e^{2\pi i 2^{n-2}\phi}|1\rangle\big)\cdots \big(|0\rangle + e^{2\pi i 2^0\phi}|1\rangle\big) \; |u\rangle \nolabel \\
& & \qquad \qquad \qquad  \qquad \qquad  = {1\over \sqrt{2^n}} \sum_{k=0}^{2^n-1}e^{(2\pi i) k \phi }|k\rangle|u\rangle.
\end{eqnarray}

Finally, we can do an \it inverse \rm QFT to the first register, taking
\begin{eqnarray}
{1\over \sqrt{2^n}} \sum_{k=0}^{2^n-1}e^{(2\pi i) k \phi }|k\rangle|u\rangle \rightarrow |\tilde \phi\rangle |u \rangle,
\end{eqnarray}
where $\tilde \phi$ is an estimator for $\phi$ when the first register is measured.  
If $n$ q-bits are used in the first register (as indicated above), then this algorithm will accurately approximate $\phi$ to 
\begin{eqnarray}
n- \bigg\lceil \log_2 \bigg(2+{1\over 2\epsilon}\bigg) \bigg\rceil \label{eq:howacc}
\end{eqnarray}
bits with probability $1-\epsilon$.  

\subsection{Grover's Quantum Search Algorithm}\label{sec:grover}

We now consider the algorithm at the heart of our application to superstring model building statistics, Grover's search algorithm  \cite{quantcom} \cite{mer} \cite{grover}.  The idea is that we start with a space of $N=2^n$ elements which we want to search through.  We index the elements in the space, which requires $n$ bits.  We are looking for elements which satisfy some specific criterion, of which assume there are $M$ in the search space.  

Next, we define what is called an \it oracle\rm, which is a function $f(x)$ which takes an $n$-bit value $x \in \{0,\ldots N-1\}$, and has the value $f(x)=0$ if $x$ is not one of the $M$ solutions, and the value $f(x)=1$ if $x$ is one of the $M$ solutions.  We will not concern ourselves with the construction of the oracle in this section or the next.  We simply assume its existence.  

The actual operation of the oracle is as follows; given an $n+1$ q-bit state prepared as $|x\rangle^{\otimes n}|y\rangle$, the oracle operator $\mathcal{O}$ makes the transformation
\begin{eqnarray}
\mathcal{O}|x\rangle^{\otimes n}|0\rangle = |x\rangle^{\otimes n} |y\oplus f(x)\rangle
\end{eqnarray}
(where $\oplus$ denotes addition mod $2$).  In other words, $\mathcal{O}$ changes the value of the last q-bit iff $x$ is a solution.  

A more useful application of this is to prepare the last q-bit in the state $|0\rangle$ and do a Hadamard transformation (\ref{eq:hadamard}).  Defining $|-\rangle \equiv {1\over \sqrt{2}}\big(|0\rangle - |1\rangle \big)$, we then have 
\begin{eqnarray}
\mathcal{O}|x\rangle^{\otimes n} |-\rangle = (-1)^{f(x)}|x\rangle |-\rangle.
\end{eqnarray}
To carry out the actual search algorithm, the computer is prepared in the state $|0\rangle^{\otimes n}|-\rangle$, and the each of the first $n$ states are acted on by Hadamard gates, producing
\begin{eqnarray}
|\psi\rangle = {1\over \sqrt{2^n}} \sum_{x=0}^{2^n-1} |x\rangle |-\rangle. \label{eq:initialstate}
\end{eqnarray}

To this state we apply the ``Grover Operator" $G$ defined as
\begin{eqnarray}
G \equiv H^{\otimes n}(2|0\rangle \langle 0| - I)H^{\otimes n} \mathcal{O} = (2|\psi\rangle \langle \psi| - \mathbb{I})\mathcal{O}.
\end{eqnarray}
Let $|\alpha\rangle$ be the normalized superposition of all states which are not solutions, $|\alpha\rangle = {1\over \sqrt{N-M}} \sum_{x\ni f(x)=0} |x\rangle$, and $|\beta\rangle$ those that are, $|\beta\rangle = {1\over \sqrt{M}} \sum_{x \ni f(x)=1} |x\rangle$.  Then, (\ref{eq:initialstate}) can be written as
\begin{eqnarray}
|\psi\rangle = \sqrt{{N-M \over M}} |\alpha\rangle + \sqrt{{M\over N}}|\beta\rangle.
\end{eqnarray}
(we have dropped $|-\rangle$ for notational simplicity).  

It is clear that $\mathcal{O}$ is a reflection in the ($|\alpha\rangle$, $|\beta\rangle$) plane about $|\alpha\rangle$, and that $2|\psi\rangle \langle \psi| - \mathbb{I}$ is a reflection in the same plane about $|\psi\rangle$.  So, $G$ is a \it rotation \rm in the ($|\alpha\rangle$, $|\beta\rangle$) plane.  If we define 
\begin{eqnarray}
\cos {\theta \over 2} \equiv \sqrt{{N-M \over N}}, \label{eq:deftheta}
\end{eqnarray}
then $k$ successive applications of $G$ will produce the state
\begin{eqnarray}
G^k |\psi\rangle = \cos\bigg({2k+1\over 2}\theta\bigg)|\alpha\rangle + \sin \bigg({2k+1\over 2}\theta \bigg)|\beta\rangle.
\end{eqnarray}

If we do approximately $k={\pi \over 4}\sqrt{N\over M}$ applications of $G$, we will have rotated the state very near $|\beta\rangle$, so that measurement of $|\psi\rangle$ will give a solution with high probability.  
So, whereas with a classical computer, $O(N/M)$ applications of the oracle are needed, only $O(\sqrt{N/M})$ are needed with the quantum computer.  

\subsection{Quantum Counting}

Finally we review the Quantum Counting algorithm \cite{quantcom}, which combines ideas from the previous sections.  Grover's algorithm required that we know both the total search space $N$, as well as the number of solutions $M$.  But, it is possible that $M$ is not known in advance.  In this case, there is a fast quantum algorithm which will calculate $M$ to high probability.  
The primary idea is to use the phase estimation algorithm of section \ref{sec:phase} to find the eigenvalue of the Grover operator $G$, from which we can calculate $\theta$ (cf. section \ref{sec:grover}), and therefore $M$ (assuming we know $N$).  
The eigenvalues of the Grover operator $G$ in the space spanned by $|\alpha\rangle$ and $|\beta\rangle$ are $e^{\pm i \theta}$.  We can therefore use the phase estimation algorithm to find an estimate for $\theta$ to whatever accuracy we wish.  Once $\theta$ is obtained, we can use (\ref{eq:deftheta}) to find $M$.  

\section{Statistics on the Free Fermionic Landscape}\label{sec:sffl}

In \cite{mr1} we found that for a given layer $L$ and total order product
\begin{eqnarray}
A \equiv \prod_{i=1}^L(N_i) - 1,
\end{eqnarray}
where $N_i$ are the respective orders of the $i=1,\ldots L$ $\vec \alpha^i$'s, the number of modular invariant input sets can be approximated by
\begin{eqnarray}
B  = \bigg(\prod_{i=1}^L \prod_{j=1}^{i-1}N_{ij}\bigg){\Gamma(23+A) \over 2^LA!(A+1)^L\Gamma(23)}. \label{eq:totalnumber}
\end{eqnarray}
Collecting statistics then consists of analyzing each model, looking for occurrences of a particular gauge group.  In other words, if we want to know what percentage of all layer $L$ and order $A$ sets contain a group $M$, we must analyze all $B$ sets.  So, given an oracle which takes a set of $\vec \alpha^i$'s and returns $1$ if $M$ is produced and $0$ otherwise, the oracle must be invoked $B$ times to get an exact statistic.  
Obviously there will be a limit on $A$ and $L$, beyond which we can no longer perform comprehensive investigations.  
Graphing (\ref{eq:totalnumber}) vs. $A$ logarithmically (at $L=1$),
\begin{center}
\includegraphics[scale=.8]{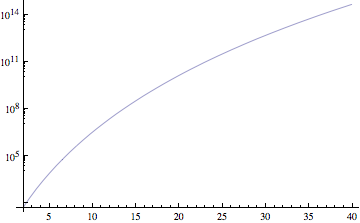}
\end{center}
we see that the computational limit will be between $20$ and $25$.  

With a quantum computer, however, we can combine Grover's algorithm (section \ref{sec:grover}) and the phase estimation algorithm (section \ref{sec:phase}) to improve our search capabilities drastically.  Equation (\ref{eq:totalnumber}) gives us a good approximation for the total number of possibilities at a given $L$ and $A$, and therefore all we need to know is the phase of the Grover operator eigenvalue for an oracle that returns $1$ for group $M$, and we can calculate the percentage of sets (at that $L$ and $A$) that will contain $M$.  

For $L$ and $A$, we can index each set using at most $\log_2 (B)$ q-bits.  So, equation (\ref{eq:howacc}) demands that the first register in our phase estimation algorithm contain 
\begin{eqnarray}
k=\log_2 B + \bigg\lceil \log_2\bigg(2+{1\over 2\epsilon}\bigg)\bigg\rceil,
\end{eqnarray}
and therefore the phase estimation will involve only $k$ invocations of the oracle.  
So, if we want to measure the eigenvalue with $99\%$ probability, our phase estimation procedure will only involve 
\begin{eqnarray}
k=\log_2 \bigg[\bigg(\prod_{i=1}^L \prod_{j=1}^{i-1}N_{ij}\bigg){\Gamma(23+A) \over 2^LA!(A+1)^L\Gamma(23)} \bigg]+ 6.
\end{eqnarray}
We can graph $k$ vs. $A$ logarithmically (at $L=1$),
\begin{center}
\includegraphics[scale=.8]{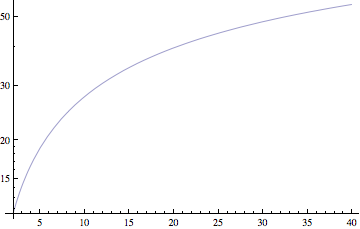}
\end{center}
which is obviously a drastic improvement over the classical approach.  We can see the improvement more clearly by graphing ${k \over B}$:
\begin{center}
\includegraphics[scale=.8]{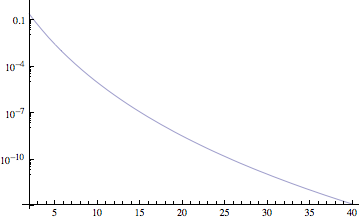}
\end{center}

\section{Conclusion}\label{sec:concl}

When Grovers search algorithm is combined with the phase finding algorithm based on the quantum Fourier transform, an enormous improvement in the depth of statistical analysis on the free fermionic heterotic string Landscape over classical computation can be obtained.  The massive parallelism made possible by the superposition of states would render such a statistical search effectively unbounded.  
However, because quantum computing is still in its infancy and no quantum computers are in operation, the purpose of this paper is merely to outline a specific application of an already existing quantum computing algorithm for a potential future use in string theory.  

\section{Acknowledgements}

Research funding leading to this manuscript was partially provided by Baylor URC grant 0301533BP.

\end{document}